\documentstyle[12pt]{article}
\def\shiftdown#1{#1\llap{\lower.04ex\hbox{#1}}}

\input epsf

\begin{document}

\author{Amand Faessler$^1$,\\ M. I. Krivoruchenko$^{1,2}$ and
B. V. Martemyanov$^{1,2}$ \\
{\small $^1${\it Institut f\"ur Theoretische Physik, Universit\"at
T\"ubingen, Auf der Morgenstelle 14 }}\\
{\small {\it D-72076 T\"ubingen, Germany}}\\
{\small $^2${\it Institute for Theoretical and Experimental Physics,
B.Cheremushkinskaya 25}}\\
{\small {\it 117259 Moscow, Russia}}}
\title{The Condensation of Dibaryons in Nuclear Matter and Its
Possible Signatures in Heavy Ion Collisions}
\date{}
\maketitle
\begin{abstract}
We consider the thermodynamics of the matter made of equal number
of neutrons and protons and of scalar dibaryons. They interact via
the exchange of scalar and vector mesons. The interaction is
taken into account in the mean field approximation. The condensation
of dibaryons in this matter and the phase transition of matter
to quark matter are considered. Possible signatures of dibaryons
in Heavy Ion Collisions are speculated on.\\


{\bf keywords}: nuclear matter, dibaryon, condensation

\end{abstract}
\newpage
\section{Introduction}

The possible existence and properties of a narrow dibaryon $d^\prime
(T=0, J^P=0^{-})$ are discussed widely in the last years both from
experimental [1,2,3] and theoretical [4,5] point of view.
Being bosons such particles could condense in nuclear matter
under proper conditions. This phenomenon happens when the
nucleon chemical potential grows up to  half of the mass
of the dibaryon. Dibaryons are weakly produced due to the reactions
\begin{equation}
\begin{array}{c}
n + p \rightarrow d^\prime + \pi^0\\
n + n \rightarrow d^\prime + \pi^-\\
p + p \rightarrow d^\prime + \pi^+
\end{array}
\end{equation}
which put them in chemical equilibrium with protons and neutrons.
At zero temperature the condensation of dibaryons in nuclear ($n = p$)
and neutron matter
was studied in Refs.[6,7,8]. For the case of $d^\prime$ the
condensation starts when the baryon number density $\rho$ grows
up to approximately $3\rho_0$ (here $\rho_0$ is the normal nuclear
density). Such conditions (zero temperatures and high densities) are
hard to imagine in terrestrial experiments. In neutron stars the
densities of the central core can be even larger but it is difficult
to see any clear signal [7] of the dibaryon condensation in neutron
stars.

High densities of nuclear matter can in principle occur in
Heavy Ion Collisions (HIC). But in that case not only the densities
but also the temperatures are high. So we should extend the consideration
of the dibaryon condensation to finite temperatures. Secondly
we should take into account the possible phase transition of
nuclear matter into quark matter.

Below we will describe our model of nuclear matter with dibaryons in
section 2, consider the thermodynamical quantities of nuclear matter
in section 3 (condensation curve), describe the model for quark matter in
section 4 (phase transition) and speculate on possible signatures of
dibaryons in HIC in the conclusions.
\section{The model of nuclear matter with dibaryons}

We consider an extension of the Walecka model [8] by including dibaryon fields
to the Lagrangian density
\begin{equation}
\label{1}
\begin{array}{c}
L=\bar \Psi (i\hat \nabla -m_N-g_\sigma \sigma -g_\omega \hat \omega )\Psi
+\frac 12(\partial _\mu \sigma )^2-\frac 12m_\sigma ^2\sigma ^2-\frac
14F_{\mu \nu }^2+\frac 12m_\omega ^2\omega _\mu ^2+ \\
(\partial _\mu -ih_\omega \omega _\mu )\varphi ^{*}(\partial _\mu +ih_\omega
\omega _\mu )\varphi -(m_D+h_\sigma \sigma )^2\varphi ^{*}\varphi .
\end{array}
\end{equation}
Here, $\Psi $ is the nucleon field, $\omega _\mu $ and $\sigma $ are the $%
\omega $- and $\sigma $-meson fields, $F_{\mu \nu }=\partial _\nu \omega_\mu
-\partial _\mu \omega _\nu $ is the field strength tensor of the vector
field; $\varphi $ is the dibaryon field, for which we assume that it is a
isoscalar-pseudoscalar field. This assumption includes
the interesting case of the  $d^{\prime }$-dibaryon.
The values $m_\omega \ $and $m_\sigma $ are the $\omega $- and $\sigma $%
-meson masses and $g_\omega $, $g_\sigma $, $h_\omega $, $h_\sigma $ are the
coupling constants of the $\omega $- and $\sigma $-mesons with nucleons ($g$%
) and dibaryons ($h$).

The field operators can be expanded into $c$-number- and operator parts:
\begin{equation}
\label{6}
\begin{array}{c}
\sigma =\sigma _c+\hat \sigma , \\
\omega _\mu =g_{\mu 0}\omega _c+\hat \omega _\mu , \\
\varphi =\varphi _c+\hat \varphi , \\
\varphi ^{*}=\varphi _c^{*}+\hat \varphi ^{*}.
\end{array}
\end{equation}
The $c$-number parts of the fields $A=$ $\sigma $, $\omega _\mu $, $\varphi $%
, and $\varphi ^{*}$ are defined as expectation values $A_c=<A>$ over the
ground state of the system. The average values of the operator parts are
zero by definition: $<\hat A>=\;0$.

The $\sigma $-meson mean field determines the effective nucleon and dibaryon
masses in the medium
\begin{equation}
\label{7}m_N^{*}=m_N+g_\sigma \sigma _c,
\end{equation}
\begin{equation}
\label{8}m_D^{*}=m_D+h_\sigma \sigma _c.
\end{equation}

The nucleon vector and scalar densities are defined by expectation values
\begin{equation}
\label{9}\rho _{NV}=<\bar \Psi \gamma _0\Psi >,
\end{equation}
\begin{equation}
\label{10}\rho _{NS}=<\bar \Psi \Psi >.
\end{equation}
The vector and scalar density of the dibaryons are defined by
\begin{equation}
\label{11} \rho _{DV}=<\varphi ^{*}i\stackrel{\leftrightarrow
}{\partial }%
_0 \varphi -2h_\omega \omega _0 \varphi ^{*}\varphi >
,
\end{equation}
\begin{equation} \label{12}\rho _{DS}=
<\varphi^{*} \varphi>.
\end{equation}

Neglecting the operator parts of the meson fields,
we get the following expressions for the meson mean fields
\begin{equation}
\label{13}\omega _c=\frac{g_\omega \rho _{NV}+h_\omega
\rho_{DV}}{m_\omega ^2}, \end{equation}
\begin{equation}
\label{14}\sigma _c=-\frac{g_\sigma \rho _{NS}+h_\sigma
2m_D^{*}\rho _{DS} }{m_\sigma ^2}.
\end{equation}

The nucleon vector and scalar densities are given by
\begin{equation}
\begin{array}{c}
\label{15}\rho _{NV}=\gamma \int \frac{d{\bf k}}{(2\pi )^3}
((exp(\frac{\sqrt{m_N^{*2}+{\bf k}^2}-\nu_N}{\theta})+1)^{-1}
-(exp(\frac{\sqrt{m_N^{*2}+{\bf k}^2}+\nu_N}{\theta})+1)^{-1}),
\end{array}
\end{equation}
\begin{equation}
\begin{array}{c}
\label{16}\rho _{NS}=\gamma \int \frac{d{\bf k}}{(2\pi
)^3}\frac{m_N^{*}}{\sqrt{m_N^{*2}+{\bf k}^2}}
((exp(\frac{\sqrt{m_N^{*2}+{\bf k}^2}-\nu_N}{\theta})+1)^{-1}
+(exp(\frac{\sqrt{m_N^{*2}+{\bf k}^2}+\nu_N}{\theta})+1)^{-1}).
\end{array}
\end{equation}
The statistical factor $\gamma =4$ $(2)$ for nuclear (neutron) matter.
Here $\nu_N$ is the nucleon chemical potential (excluding
electrostatic energy of nucleon in the vector field )
$\nu_N=\mu _N - g_\omega \omega^c$.

The dibaryon vector and scalar densities are given by
\begin{equation}
\begin{array}{c}
\label{17}\rho _{DV}=\gamma_D \int \frac{d{\bf k}}{(2\pi )^3}
((exp(\frac{\sqrt{m_D^{*2}+{\bf k}^2}-\nu_D}{\theta})-1)^{-1}
-(exp(\frac{\sqrt{m_D^{*2}+{\bf k}^2}+\nu_D}{\theta})-1)^{-1})+\\
+\rho_{DV}^0=\\
=\rho_{DV}^\prime+
\rho_{DV}^0,
\end{array}
\end{equation}
\begin{equation}
\begin{array}{c}
\label{18}\rho _{DS}=\gamma_D \int \frac{d{\bf k}}{(2\pi
)^3}\frac{1}{2\sqrt{
m_D^{*2}+{\bf k}^2}}
((exp(\frac{\sqrt{m_D^{*2}+{\bf k}^2}-\nu_D}{\theta})-1)^{-1}
+(exp(\frac{\sqrt{m_D^{*2}+{\bf k}^2}+\nu_D}{\theta})-1)^{-1})+\\
+\frac{1}{2m_D^{*}}\rho_{DV}^0=\\
=\rho_{DS}^\prime+\rho_{DS}^0.
\end{array}
\end{equation}
Here $\gamma_D$ is the dibaryon statistical factor (1 for $d^\prime$),
$\nu_D$ is the
dibaryon chemical potential (excluding
electrostatic energy of dibaryon in the vector field)
$\nu_D=\mu _D - h_\omega \omega^c$, $\nu_D \le m_D^*$,
$\rho_{DV}^0$ is the density of dibaryons in the condensate
that is formed when $\nu_D = m_D^*$.
The chemical equilibrium between nucleons and dibaryons means
$\nu_D = 2\nu_N$ (we consider the case when electrostatic energy is
not changed in the process of transition of two nucleons to dibaryon
($h_\omega = 2g_\omega$)).

 The self-consistency equations
have the form \begin{equation} \label{19}m_N^{*}=m_N-\frac{g_\sigma
}{m_\sigma ^2}(g_\sigma \rho _{NS}+h_\sigma (2m_D^{*}\rho_{DS}^\prime+
\rho _{DV}^0)),
\end{equation}
\begin{equation} \label{20}m_D^{*}=m_D-\frac{h_\sigma
}{m_\sigma ^2}(g_\sigma \rho _{NS}+h_\sigma (2m_D^{*}\rho_{DS}^\prime+
\rho _{DV}^0)).
\end{equation}
The total baryon number
density equals $\rho _{TV}=\rho _{NV}+2\rho _{DV}.$

We use here the following procedure of solving the self-consistency
equations. We start with fixing the effective neucleon mass $m_N^*$.
The value of effective dibaryon mass then simply follows from eqs.
(\ref{19},\ref{20})
\begin{equation}
\label{21}
m_D^* = m_D - \frac{h_\sigma}{g_\sigma}(m_N -m_N^*).
\end{equation}
Further we assume that the condensation takes place and
find the nucleon chemical potential: $\nu_N = m_D^*/2$.
Then we find the condensate dibaryon density $\rho_{DV}^0$ from eq.(\ref{19}).
If it is positive the self-consistency equation is solved for a given
point. If it is negative we take $\rho_{DV}^0 = 0$ and scan $\nu_N = \nu_D/2$
from $0$ to $m_D^*/2$ to make the eq.(\ref{19}) true. After this procedure
all the parameters of the matter are defined and we are ready to calculate
various thermodynamical quantities.
\section{Thermodynamical quantities of nuclear matter. Condensation curve}

The energy density $\varepsilon =<T_{00}>$ given by average value of the $%
T_{00}$ component has the form
\begin{equation}
\begin{array}{c}
\label{26}\varepsilon =\gamma \int\frac{d{\bf k}}{(2\pi )^3}%
((\sqrt{m_N^{*2}+{\bf k}^2}+g_\omega \omega _c)
(exp(\frac{\sqrt{m_N^{*2}+{\bf k}^2}-\nu_N}{\theta})+1)^{-1}\\
+(\sqrt{m_N^{*2}+{\bf k}^2}-g_\omega \omega _c)
(exp(\frac{\sqrt{m_N^{*2}+{\bf k}^2}+\nu_N}{\theta})+1)^{-1})\\
+\gamma_D \int\frac{d{\bf k}}{(2\pi )^3}%
((\sqrt{m_D^{*2}+{\bf k}^2}+h_\omega \omega _c)
(exp(\frac{\sqrt{m_D^{*2}+{\bf k}^2}-\nu_D}{\theta})-1)^{-1}\\
+(\sqrt{m_D^{*2}+{\bf k}^2}-h_\omega \omega _c)
(exp(\frac{\sqrt{m_D^{*2}+{\bf k}^2}+\nu_D}{\theta})-1)^{-1})\\
+(m_D^{*}+h_\omega \omega _c)\rho
_{DV}^0+\frac 12m_\sigma ^2\sigma _c^2-\frac 12m_\omega ^2\omega _c^2.
\end{array}
\end{equation}
The last two terms are here the
contributions of the classical $\omega $- and $\sigma $-meson fields to the
energy density.

The hydrostatic pressure $p=-\frac 13<T_{ii}>$ has the form
\begin{equation}
\begin{array}{c}
\label{27}p =\frac 13\gamma \int\frac{d{\bf k}}{(2\pi )^3}%
(\frac{{\bf k}^2}{\sqrt{m_N^{*2}+{\bf k}^2}}
(exp(\frac{\sqrt{m_N^{*2}+{\bf k}^2}-\nu_N}{\theta})+1)^{-1}\\
+\frac{{\bf k}^2}{\sqrt{m_N^{*2}+{\bf k}^2}}
(exp(\frac{\sqrt{m_N^{*2}+{\bf k}^2}+\nu_N}{\theta})+1)^{-1})\\
+\frac 13\gamma_D \int\frac{d{\bf k}}{(2\pi )^3}%
(\frac{{\bf k}^2}{\sqrt{m_D^{*2}+{\bf k}^2}}
(exp(\frac{\sqrt{m_D^{*2}+{\bf k}^2}-\nu_D}{\theta})-1)^{-1}\\
+\frac{{\bf k}^2}{\sqrt{m_D^{*2}+{\bf k}^2}}
(exp(\frac{\sqrt{m_D^{*2}+{\bf k}^2}+\nu_D}{\theta})-1)^{-1})\\
-\frac 12m_\sigma ^2\sigma _c^2+\frac 12m_\omega ^2\omega _c^2.
\end{array}
\end{equation}
Because dibaryons in the condensate are at rest they do not contribute to
the pressure.

Now we are in a position to present the results for the thermodynamical
quantities of nuclear matter. For our purpose the most interesting
thing is the condensation curve i.e. the curve on the temperature-
baryon density plane where the condensation of dibaryons starts
either at increasing the baryon density or at lowering the temperature.
The calculations are made for the standart values of RMF model
parameters[9]. The coupling constants of the dibaryon $d^\prime(2060)$
to $\sigma -$ and $\omega -$ mesons were taken to be[8]
$$h_\omega = 2g_\omega~~~~~~~~~~~~~~~h_\sigma = 1.6g_\sigma$$
The results are shown on the Fig.1 (condensation curve).
It means e.g. that at zero temperature ($T = 0$) one has to
compress to $\rho_{TV} = 0.54 fm^{-3}$ ($\approx 2.7$ of saturation
density) to obtain the transition to the dibaryon condensate
if the dibaryon mass is $m_{d^\prime} = 2060 MeV $.
Also we show on Fig.1 the so called critical curve. It is the
curve that shows the boundary of applicability of our model.
Precisely it is the place where the effective mass of the nucleon
goes to zero. On the right of the critical curve our model cannot
be applied. We hope that the more elaborated model will move
this curve to the right in the unphysical region where the
nuclear matter transforms to the quark matter for example.
And finally we show on Fig.1 the phase transition curve i.e.
the place where the nuclear matter (with or without condensed
dibaryons) transforms to quark matter. Now we are coming to the
description of this curve.
\section{The model of quark matter. Phase transition}

The nuclear matter we are considering is isotopically symmetric
($\rho_p = \rho_n$)
and has no net strangeness($\rho_s = 0$). At the beginning of the
phase transition to quark matter
isospin and strangeness are conserved. Therefore we will
compare the thermodynamical quantities of nuclear matter with those
of quark matter consisting of equal number of $u$ and $d$ quarks and
no net strange quarks. The reliable temperatures in our consideration
(we are interested in nuclear matter with condensed dibaryons) will
be below $50 MeV$. For such temperatures the contribution of pions,
kaons etc. to the thermodynamical quantities of nuclear matter and
the contribution of strange quarks, gluons etc. to the thermodynamical
quantities of quark matter can be neglected. Then the equation of
state (EOS) of nuclear matter $p_N(\mu_N,T)$ is described by eq.
(\ref{27}) and the equation of state of quark matter
$p_q(\mu_q,T)$ is the following one[11]
\begin{equation}
\label{31}p_q=(\frac{1}{2\pi^2}\mu_q^4 + \mu_q^2T^2)(1-\frac{2\alpha_c}{\pi})
+ \frac{16}{3\pi}(\frac{\alpha_c}{\pi})^\frac 32 \mu_q^3T -B.
\end{equation}
Here $\alpha_c$ is the constant of gluonic interaction of quarks.
The first term in eq.(\ref{31}) is the pressure of massless $u$ and $d$
quarks,
the second term is the plasmon contribution to the pressure and
the third term is the vaccuum pressure  that is responsible for
confinement (MIT bag model[11]).
The constants $\alpha_c$ and $B$ were taken to be
$$
\alpha_c = 0.6~~~~~~~~~~~~~~~~~~~B = 96~MeV/fm^3
$$
that ensures the stability of massive neutron stars against
the transition to strange stars[12].

The phase transition of nuclear matter to quark matter starts
when the pressure of quark matter becomes equal to the pressure
of nuclear matter at equal baryon chemical potentials and
temperatures (Gibbs criterion, here we assume that the phase
transition is of the first order)
$$p_N(\mu_N,T) = p_q(\mu_q,T)~~~~~~~~~~~\mu_N = 3\mu_q
$$
The typical dependences of $p_N$ and $p_q$ on $\mu_N$ for fixed $T=50MeV$
are shown on Fig.2. The point of crossing of lines is the point of
phase transition at given temperature. The full phase transition curve
is shown on Fig.1. As it follows from Fig.1 at temperatures above
$36 MeV$ the dibaryon condensation does not occur: increasing the
density for fixed temperatures above $36 MeV$ we come to the quark
matter rather than to dibaryon condensation. The region on the ($T, \rho_{TV}$)
plot where dibaryon condensate is formed ($p,n,d^\prime$ in Fig.1)
lies therefore below
$36 MeV$ for temperatures and
above $0.54~1/fm^3$ for densities (2.7 of normal nuclear density).

In conclusion let us speculate on the possible signatures
of dibaryon condensate formation in HIC.

\section{Conclusion}

The accessible temperatures and densities of nuclear matter in HIC are
not very well known. Fig.3 shows our calculations of the
temperature of compressed nuclear matter formed in HIC as a function of the
density of
compressed matter (different energies per nucleon of colliding ions).
We have used the EOS of nuclear matter with dibaryons and a one-
dimensional shock wave model[13] (Rankin-Hugoniot-Taub equation).
The compression curve lies above
the region of condensation (in $"p,n"$ or $"Q"$)
shown on Fig.1. On the first sight we should conclude that the
dibaryon condensate never forms in HIC. But the compressed matter being formed
evolves further and its evolution is governed by two effects: by the
cooling due to emission of particles (mainly pions) and by the
decompression (that also results in the cooling). If the radiation
cooling prevails over the cooling due to the decompression,
the matter could evolve
into the region of the formation of the dibaryon condensate ($p,n,d^\prime$
 in Fig.1). If so, the
significant part of nucleons will transform to dibaryons (up to $70\%$).
The dibaryons could survive in the process of decompression
($\Gamma_{d^{\prime}} \approx 0.5 MeV$) and then decay as free
particles in one of the following ways
\begin{equation}
\begin{array}{c}
d^\prime \rightarrow nn\pi^+\\
d^\prime \rightarrow np\pi^0\\
d^\prime \rightarrow pp\pi^-.
\end{array}
\end{equation}
The pions in these decays have the spectrum with the characteristic kinetic
energy about $40-50 MeV$, if the dibaryon has a mass about $2060 MeV$.
Then these pions could form an excess
in the temperature spectrum of all pions produced in HIC. In order
to these pions could be observed they should not be thermalised in
the nuclear matter like the other pions (direct ones, pions from
different other resonances not considered in this paper). The
principal possibility of this is due to the relatively long lifetime
of $d^\prime$-dibaryon - it could decay after the decay of nuclear
matter.

So,
we conclude, there is a region of temperatures and densities where
the dibaryon condensate forms in nuclear matter. If formed in
HIC, dibaryons could give a signature in  form of $40-50 MeV$ kinetic energy
pions -the products of their decays.

The authors acknowledge the discussions with Drs. D.Kosov and L.Sehn.
Two of us (M.I.K and B.V.M.) are grateful to the Institute for
Theoretical Physics of University of Tuebingen for hospitality
and financial support. This work was supported by the Deutsche
Forschungsgemeinschaft under contract $N^0$ FA67/20-1.

{\bf Figure captions}\\
{\bf Fig.1} The condensation, phase transition and critical curves
on the temperature($T$) - baryon number density($\rho_{TV}$) plot.
The regions of nuclear matter($n,p$), nuclear matter with condensed
dibaryons($n,p,d^\prime$) and quark matter($Q$) are marked. The critical curve
shows the boundary of applicability of our model.\\
{\bf Fig.2} The example of nuclear and quark matter equations of state.
The pressures of nuclear matter($p_N$) and quark matter($p_q$) are
shown as the functions of baryon chemical potential($\mu_N$) at
the temperature $T = 50~MeV$. The crossing point is the point of
phase transition from nuclear matter to quark matter.\\
{\bf Fig.3}  The temperature of the shock wave($T$) as a function of
baryon density ($\rho $) in ion collisions for
mean field model of nuclear matter with dibaryons.


\end{document}